\documentclass[twocolumn,amsmath,amssymb,superscriptaddress,preprintnumbers,showpacs]{revtex4}

\usepackage{ulem}
\usepackage{graphics}
\usepackage{dcolumn}
\usepackage[usenames]{color}    % textcolor
\usepackage{ulem}               % underlining and crossing of text
\usepackage{bm}
\usepackage{amsmath}
\usepackage{amssymb}
\usepackage{natbib}
\usepackage{epsfig}

\RequirePackage{xspace}

\newif\ifcom
\newif\ifsec
\newif\iffig
\newif\ifgraph

\figtrue                     %
\comtrue               %
\graphtrue                   %
\begin{document}

\title{Dendritic flux avalanches in Niobium single crystal near critical temperature}

\author{M.~Gr\"{u}nzweig}\email{matthias.gruenzweig1082@gmail.com}
\affiliation{Physikalisches Institut -- Experimentalphysik II and Center for Collective Quantum Phenomena in LISA$^+$, Universit\"{a}t T\"{u}bingen, Auf der Morgenstelle 14, 72076 T\"{u}bingen, Germany}

\author{P.~Mikheenko}
\affiliation{University of Oslo, Department of Physics, Blindern 0316 Oslo, Norway}

\author{C.~Gr\"{u}nzweig}
\affiliation{Paul Scherrer Institut, Neutron Imaging and Activation Group, CH-5232 Villigen, Switzerland}

\author{S.~M\"{u}hlbauer}
\affiliation{Technische Universit\"{a}t M\"{u}nchen, Forschungsneutronenquelle Heinz Maier-Leibnitz (FRM II), D-85748 Garching, Germany}

\author{R.~Kleiner}
\author{D.~Koelle}
\affiliation{Physikalisches Institut -- Experimentalphysik II and Center for Collective Quantum Phenomena in LISA$^+$, Universit\"{a}t T\"{u}bingen, Auf der Morgenstelle 14, 72076 T\"{u}bingen, Germany}

\author{T.~H.~Johansen}
\affiliation{University of Oslo, Department of Physics, Blindern 0316 Oslo, Norway}

\date{\today}

\begin{abstract}

We report on the observation of dendritic flux avalanches in a large Niobium single crystal. In contrast to avalanches observed in thin films, they appear only in a very narrow temperature interval of about a tenth of a Kelvin near the critical temperature of Nb. At a fixed temperature, we find two sets of dendritic structures, which differ by the magnetic field required for their formation and by the maximum distance the dendrites penetrate into the sample. The effect is caused by dendritic flux penetration into thin superconducting surface layers formed in the single crystal close to the critical temperature.

\end{abstract}


\maketitle

\section{Introduction}

\label{sec:Introduction}
Avalanche dynamics plays an important role in many fields: in plant growth, developmental biology, molecular dynamics, cellular automata, evolution, etc. \cite{Paczuski96, Meinhardt76}. In type-II superconductors avalanches are observed in the motion of magnetic flux. These avalanches develop on the background of a critical-state model which was proposed by Charles P. Bean in 1962. It is a state with spatially homogeneous (critical) current density determined by pinning of magnetic flux \cite{Bean64}. In many thin film superconductors this state is suddenly destroyed by the avalanches. Such avalanches were observed in thin films of Nb, Pb, Nb$_3$Sn, NbN, MgB$_2$, or YNi$_2$B$_2$C \cite{Welling04, Qviller10, Qviller12, Duran95, Menghini05, Rudnev03, Rudnev05, Wimbush04} at temperatures below a threshold temperature. The avalanches appear due to the following fundamental reasons: (i) the motion of magnetic flux (mainly in form of vortices) releases energy, and increases the local temperature, and (ii) the increase of temperature reduces flux pinning and facilitates further flux motion \cite{Prozorov06, Denisov06, Mints81, Aranson05, Altshuler04, Rakhmanov04}. The avalanches are frequently seen in the form of dendritic (branching) structures. From an applications point of view, the study of thermomagnetic avalanches is of significant importance since such events can directly affect the performance of superconducting devices. Recently, it was reported that dendritic avalanches create voltage pulses of magnitudes on the order of 1 volt during a time span of 50 ns \cite{Mikheenko13}. They also can provide a test scenario for the theory of nonequilibrium dynamics in superconductors \cite{Altshuler04}. Nb single crystals as a type-II superconductor with relatively small Ginzburg-Landau parameter $\kappa \approx 0.74$ are also considered as a model system for systematic studies of vortex matter \cite{Muehlbauer09, Laver09, Muehlbauer11}. 

In this paper we report on the observation of completely unexpectedly dendritic flux avalanches in a bulk Nb single crystal using magneto-optical imaging (MOI). These avalanches appear only in a very narrow temperature window of about a tenth of a Kelvin near the critical temperature ($T_c$). The appearance of dendritic avalanches in bulk samples and at high temperature is a completely new phenomenon. Previously, dendritic avalanches in Nb thin films were observed at low temperatures only. They were found to have a complex morphology, starting as quasi-one-dimensional structures at $T/T_c \approx 0.35$, developing some branching and finally resulting in highly branched, "sea-weed-like" structures when approaching $T/T_c \approx 0.65$. For temperatures above $T/T_c \approx 0.65$ the magnetic flux penetrates into films smoothly with fronts well described by the critical-state model \cite{Duran95, Denisov2006}. 

One of the most significant features of dendritic avalanches caused by thermomagnetic instability is the irreproducibility of their dendritic pattern, which indicates that the branching points are not directly related to the pinning landscape or non-uniformities present in the sample \cite{Qviller12, Denisov06, Vestgarden11}. Also, features like the fast flux dynamics, the enhanced branching at higher temperatures, the preferred locations for nucleation, and the existence of a threshold field are typical for avalanches. In the investigated Nb single crystal at least two of these features are seen: a preferred location for nucleation, and the irreproducibility of dendritic pattern. Even more intriguing, two sets of dendrites were observed. One set shows a strong MOI contrast and has a high threshold field. The other set shows a weak MOI contrast without threshold field.

The paper is organized as follows. After this introductory Sec.~\ref{sec:Introduction} we describe the characterization of the sample and the basic experimental MOI-arrangement in Sec.~\ref{sec:Experimental details}. Section~\ref{sec:Results} presents our results, with a discussion of our observations given in Sec.~\ref{sec:Discussion}. Finally, Sec.~\ref{sec:Conclusions} concludes the paper.

\section{Experimental details}

\label{sec:Experimental details}
Our MOI experiments and measurements of $T_c$ have been performed on a pure Niobium bulk single crystal ($\kappa \approx 0.74$) cut from a large single crystal at zero field cooling (ZFC). It is a $10 \times 10$ mm$^2$ platelet with a constant thickness of 2 mm. The normal vector of the sample coincides with a [110] direction. The magnetic field was applied parallel to the normal vector of the sample leading to a demagnetization factor close to unity. The high crystallographic quality of the sample corresponds to a \textit{RRR} of $\approx$ 150. 

\begin{figure}[h]
	\centering
		\includegraphics[width=0.37\textwidth]{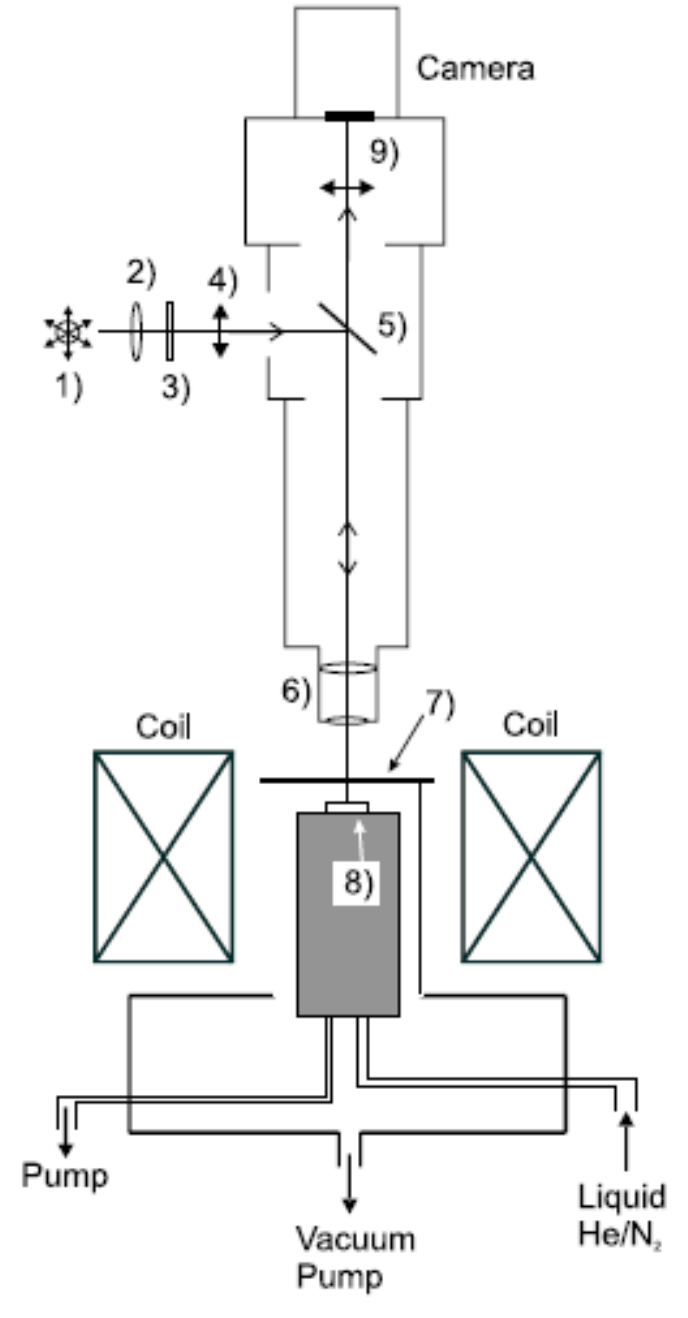}
	\caption{A sketch of the basic experimental arrangement for magneto-optical imaging of magnetic flux patterns in superconductors. The polarized light microscope has the following components: (1) stabilized light source, (2) collimator, (3) edge filter for infrared suppression, (4) polarizer, (5) semi-reflecting mirror, (6) objective, (7) covering glass of the cryostat (Suprasil), (8) sample covered with a magneto-optical layer (MOL) and (9) analyzer \cite{Jooss02}.}
	\label{fig:figure0}
\end{figure}

A schematic drawing of the MOI experimental set-up is given in Fig.~\ref{fig:figure0}.
%
%%%%%%%%Figure%%%%%%%%%%%%%%%%%%%%%%%%%%%%%%%%%%%%%%%%%%%%%
%%%%%%%%%%%%%%%%%%%%%%%%%%%%%%%%%%%%%%%%%%%%%%%%%%%%%%%%%%%%%
Magneto-optical experiments require a polarized light microscope, an optical cryostat, an electromagnetic coil for the generation of magnetic fields and an image recording system such as a camera, a video system or for digital image processing a CCD camera. The polarized light microscope consists of a stabilized light source with sufficient power, a polarizer, an analyzer and optical components to project the image plane of the magneto-optical layer (MOL) into the image recording system with various magnifications. These components, especially the objective lenses, should have small Verdet constants to avoid disturbing Faraday rotations due to magnetic stray fields of the coils. In addition, they should not depolarize the polarized light beam. Generally, magneto-optical measurements are performed in reflection mode. Due to the strong absorption of the high temperature superconductors in the frequency range of visible light, it is necessary to use mirror layers in order to obtain a better reflectivity \cite{Jooss02}.

In the MOI experiments, for direct visualization of the normal component of magnetic induction on the sample surface, a magneto-optical indicator film (Bi-doped iron garnets with in-plane magnetization) is placed on the top of the superconductor giving the incoming polarized light a Faraday rotation depending on the local magnetic field. A typical fieldsensing indicator film consists of a $\approx 500$ $\mu$m thick Gd$_3$Ga$_5$O$_{12}$ substrate, a $\approx 2$ $\mu$m thick doped iron garnet film and a reflecting aluminium layer with thicknesses of $\approx 100$ nm. After being reflected and passed through the crossed analyzer, the light produces an image which is a direct map of the distribution of the perpendicular component of the field. The MOI experiments were conducted at the University of Oslo. The images were captured with a Retiga Exi CCD camera and the videos were made with a Vision Research Phantom V 210 camera. The light was supplied by a mercury lamp and an objective lens with 2.5$\times$ magnification was used providing a field of view of 5.85 mm $\times$ 7.61 mm. The sample was mounted in an Oxford Microstat continuous Helium flow cryostat and a coil provided the out-of-plane magnetic field. This setup reaches the limits of lowest \textit{T} of 3.7 K and highest \textit{B} of 85 mT. Due to the finite thermal resistance between the cold finger and the sample, only the temperature relative to $T_c$, in which the MOI signal disappears, is quoted. When recalculated to absolute value, it is found that it remains stable during a sequence within a few mK.

\section{Results}

\label{sec:Results}
\begin{figure}[h]
	\centering
		\includegraphics[width=0.45\textwidth]{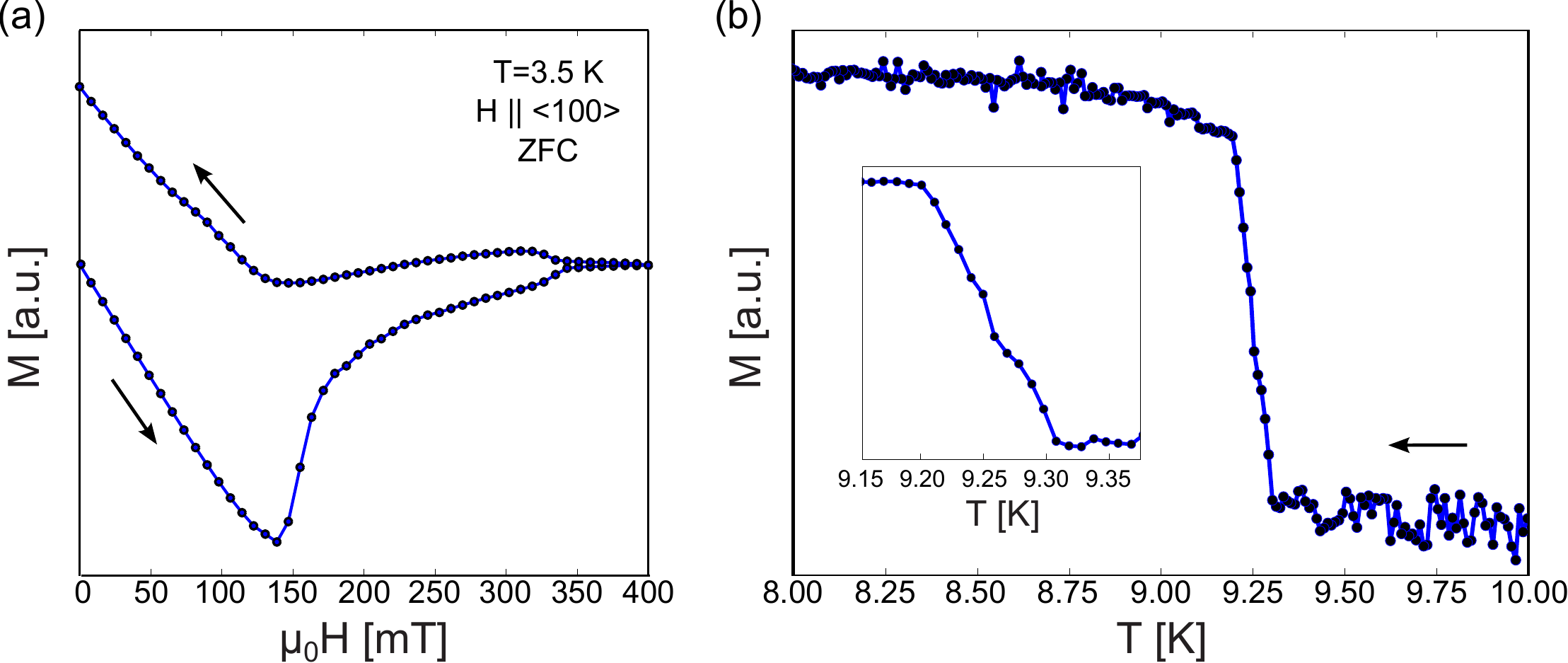}
	\caption{(Color online) (a) Hysteresis loop $M(H)$ of the sample. The large hysteresis shows that the crystal contains strong pinning centres. (b) Determination of $T_c$ from a measurement of temperature vs magnetization. The inset in (b) shows a magnified view close to the superconducting transition temperature $T_c$ of 9.25 K as taken in the middle of the transition.}
	\label{fig:figure1}
\end{figure}

The magnetic properties obtained by DC SQUID magnetometry and the determination of $T_c$ are summarized in Fig.~\ref{fig:figure1}.
%
%%%%%%%%Figure%%%%%%%%%%%%%%%%%%%%%%%%%%%%%%%%%%%%%%%%%%%%%
%%%%%%%%%%%%%%%%%%%%%%%%%%%%%%%%%%%%%%%%%%%%%%%%%%%%%%%%%%%%%
The large hysteresis seen in the magnetization \textit{M} vs magnetic field \textit{H} measurements (Fig. 1(a)) shows that the crystal contains strong pinning centers. The $T_c$, as determined by the \textit{M(T)} curve (Fig. 1(b)), is 9.25 K. This value agrees well with $T_c=9.26 \pm 0.1$ K reported by Finnemore et al. \cite{Finnemore66} for high purity Nb (residual resistance ratio $RRR=1600 \pm 400$).
 
\begin{figure}[ht]
	\centering
		\includegraphics[width=0.45\textwidth]{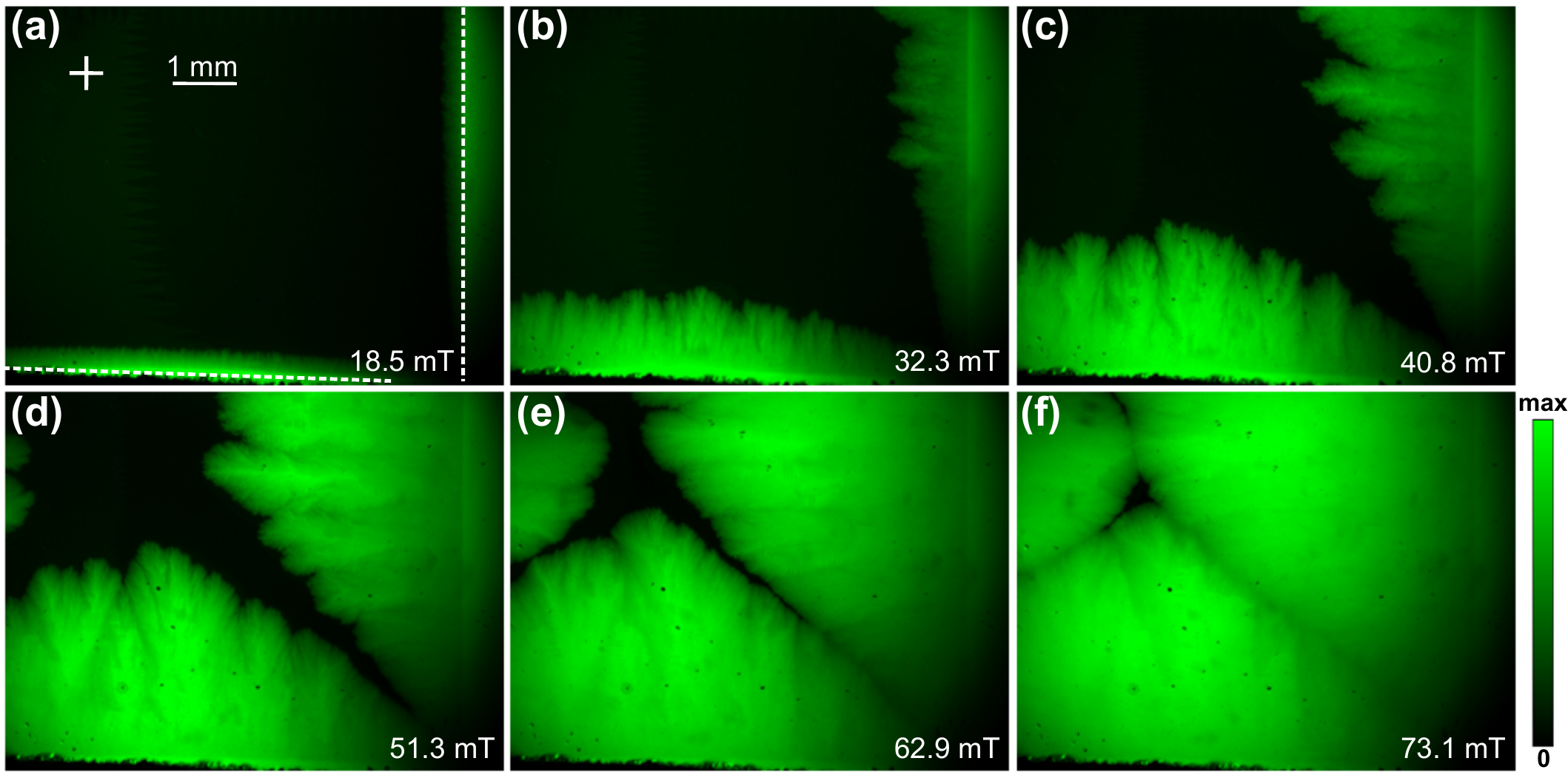}
	\caption{(Color online) MO images of flux penetration into the bulk single crystal at $T/T_c \approx 0.53$. The image brightness represents the flux density. Images (a)-(f) are taken after zero-field-cooling at different applied fields perpendicular to the sample. The flux penetrates into the bulk sample smoothly with fronts well described by the critical-state model. In (a) the sample edges are indicated by white dashed lines and the center of the sample is indicated by a white cross.}
	\label{fig:figure2}
\end{figure}

For the MOI experiments the magnetic field was applied perpendicular to the platelet-shaped crystal. In Fig.~\ref{fig:figure2}, the MO images show flux penetration into the initially zero-field-cooled sample for different magnetic fields ((a)-(f)) from 18.5 mT to 73.1 mT at $T/T_c \approx 0.53$. 
%
%%%%%%%%Figure%%%%%%%%%%%%%%%%%%%%%%%%%%%%%%%%%%%%%%%%%%%%%
%%%%%%%%%%%%%%%%%%%%%%%%%%%%%%%%%%%%%%%%%%%%%%%%%%%%%%%%%%%%%
The data were recorded using a CCD camera and are presented as original images. Black color indicates zero flux while regions containing vortices appear in green (gray). With increasing applied magnetic field, the flux penetrates from the edges deeper inside the Niobium crystal, exhibiting a relatively smooth distribution of vortices. Only the corners and the central part remain flux free at low fields. In this regime the sample behaves in agreement with the critical state model.
 
\begin{figure}[ht]
	\centering
		\includegraphics[width=0.45\textwidth]{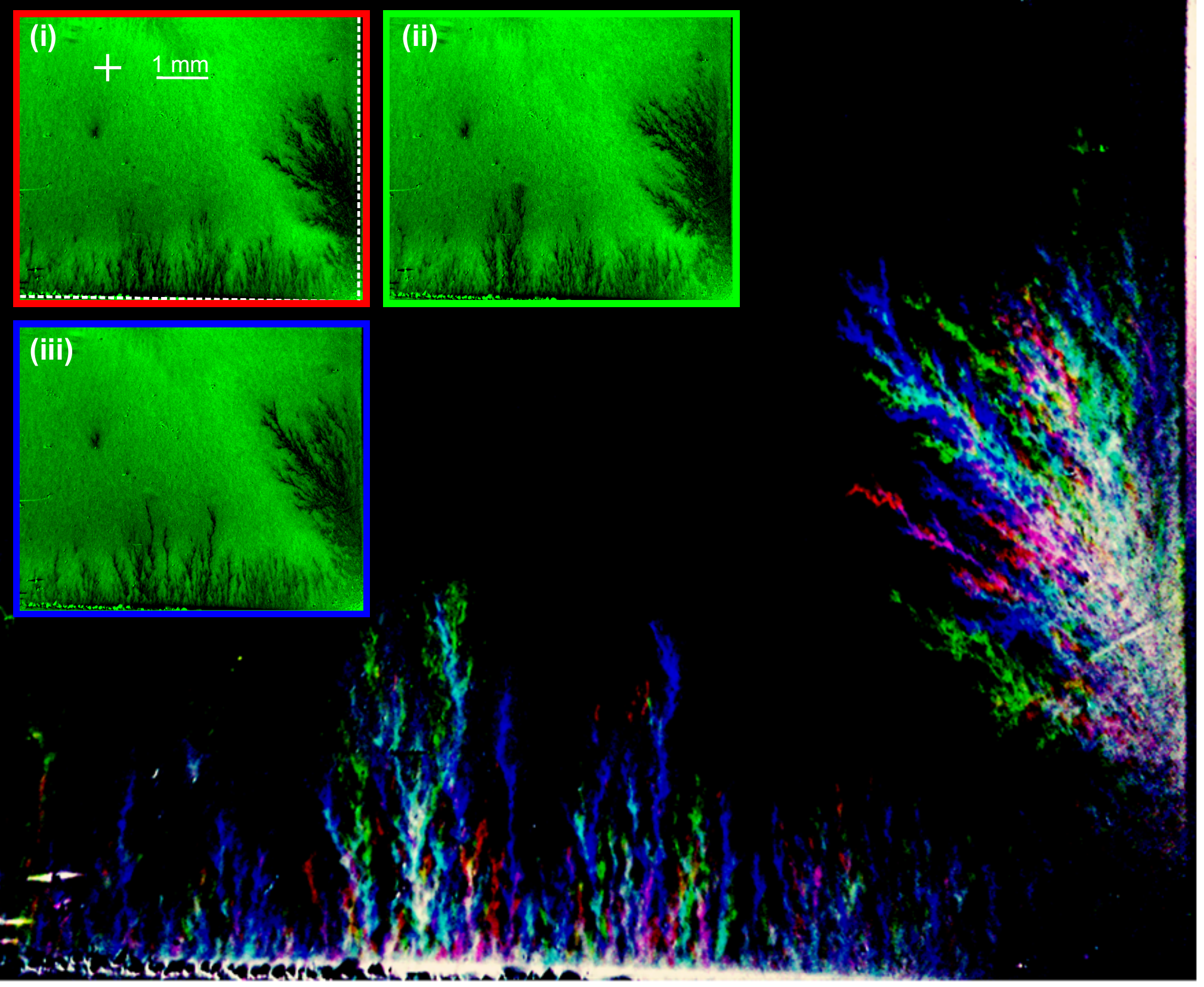}
	\caption{(Color online) Irreversibility of the dendritic structures near $T_c$. Superposition of three MO images taken in different experimental runs as shown in the insets being encoded in red (i), green (ii), and blue (iii). The dendritic flux penetration takes place at $T/T_c \approx 0.986$ after decreasing the magnetic field from $B_a= 0.68$ mT to 0. Well separated colours show that the dendritic avalanches are irreproducible. The images in the inset are subtracted from the image taken slightly above $T_c$ at $B_a=0$. The sample edges are indicated by white dashed lines and the centre of the sample is indicated by a white cross in image (i).}
	\label{fig:figure3}
\end{figure}

Above $T/T_c \approx 0.985$ magnetic flux starts penetrating in form of dendritic avalanches. To investigate the irreproducibility of the dendrites, different experimental runs under the same conditions were performed. The results are shown in Fig.~\ref{fig:figure3}.
%
%%%%%%%%Figure%%%%%%%%%%%%%%%%%%%%%%%%%%%%%%%%%%%%%%%%%%%%%
%%%%%%%%%%%%%%%%%%%%%%%%%%%%%%%%%%%%%%%%%%%%%%%%%%%%%%%%%%%%%
The insets (i)-(iii) were recorded at $T/T_c \approx 0.986$ after sweeping the applied field to 0.68 mT, which generated an almost homogeneous distribution of magnetic flux in the sample, and then abruptly turning off the field. For the inset images the contrast has been enhanced by subtracting an image taken slightly above $T_c$ at $B_a=0$. All inset images clearly show dark dendritic structures resulting from antivortices, which entered the crystal. Note that single flux quanta are not resolved in the images. In addition to the black dendrites, the images also show bright regions of positive flux which extend into the center of the crystal and, as will be shown below, are also dendritic.  Figure~\ref{fig:figure3} shows in different colors (gray scales) the superposition of the dark dendritic structures from the inset images. One can see that in each run the dendrites follow different paths, which is a typical signature of dendritic avalanches \cite{Qviller12, Denisov06, Vestgarden11}. If, by contrast, the magnetic flux follows non-uniformities, these patterns would have been reproducible. The dendritic behavior seen in Fig.~\ref{fig:figure3} is observed up to $T/T_c \approx 0.993$. For higher temperatures dendritic avalanches are still present, but behave differently.

\begin{figure}[ht]
	\centering
		\includegraphics[width=0.45\textwidth]{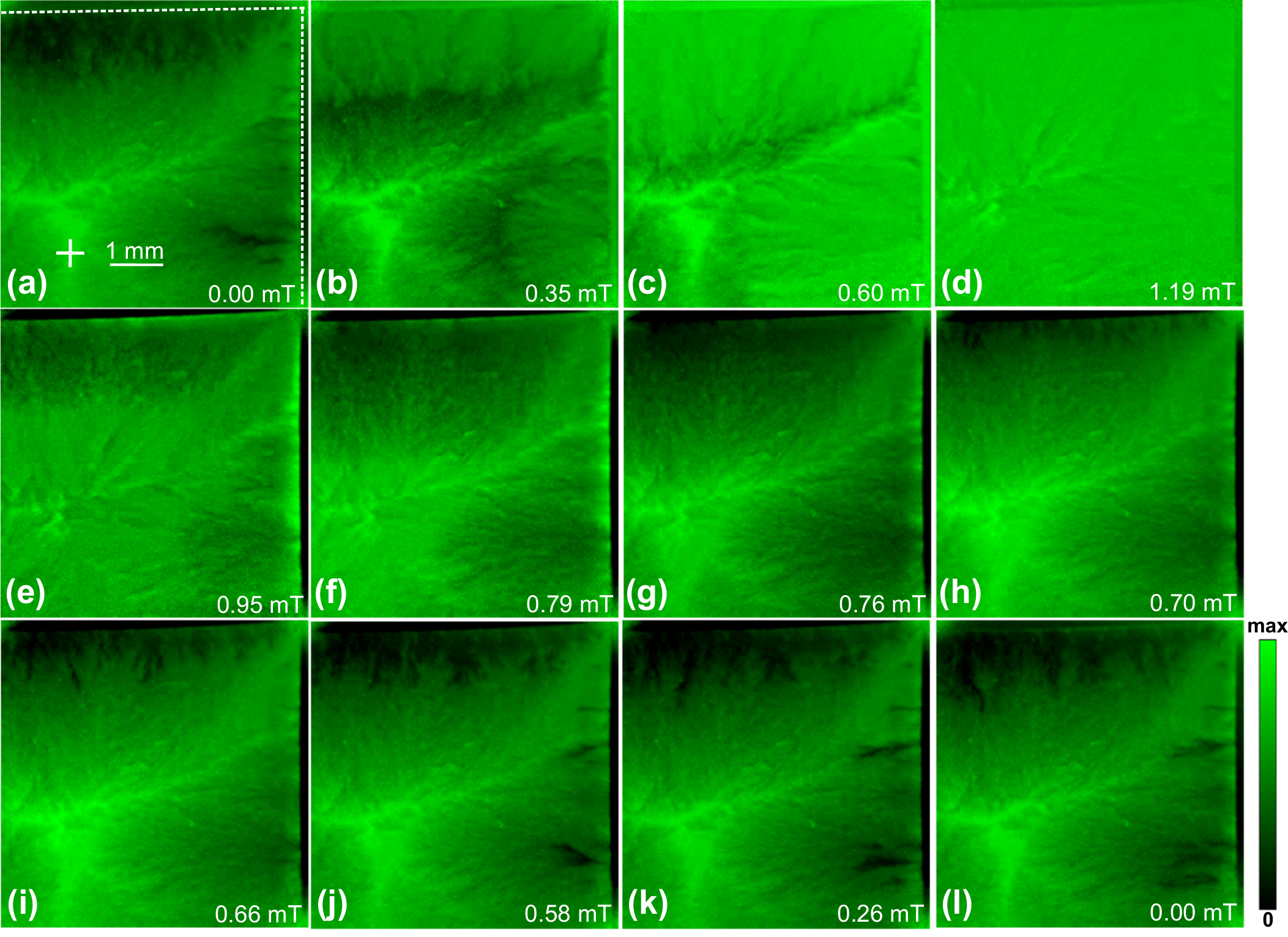}
	\caption{(Color online) MO images of dendritic flux penetration into a Nb bulk single crystal at $T/T_c \approx 0.994$. The applied magnetic field is oriented perpendicular to the sample and varied sinusoidally between 0 and 1.19 mT. Images (a)-(l) have been taken within one period of the sweep. All images are subtracted from an image taken at $T/T_c \approx 0.994$ and $B_a=1.19$ mT prior to the sweeping sequence. A video of the whole scan is shown in the supplemetary information (video Fig. 5). In (a), the sample edges are indicated by white dashed lines and the centre of the sample is marked by a white cross.}
	\label{fig:figure4}
\end{figure}

In the following experiments the applied field was changed sinusoidally between 0 and 1.19 mT with a period of 16 seconds at a fixed temperature. The data were collected using an ultra-fast CCD camera with a sampling rate of 100 frames per second leading to an exposure time of 10 ms. Figure~\ref{fig:figure4} shows a sequence of frames recorded at $T/T_c \approx 0.994$, during one period. 
%
%%%%%%%%Figure%%%%%%%%%%%%%%%%%%%%%%%%%%%%%%%%%%%%%%%%%%%%%
%%%%%%%%%%%%%%%%%%%%%%%%%%%%%%%%%%%%%%%%%%%%%%%%%%%%%%%%%%%%%
The images (a)-(d) were taken at increasing magnetic field, while images (e)-(l) are taken at decreasing field. Figure~\ref{fig:figure4}(a), recorded at $B_a=0$, contains magnetic flux present from previous field cycles. In this figure, one can see magnetic flux of high density (bright areas) in the center of the image, while near the upper and right edge of the sample dark (low magnetic flux) dendrites are visible. When the applied field is increased, bright dendrites enter the crystal. As seen in Figs.~\ref{fig:figure4}(b)-(d), these dendrites grow with field until at 1.19 mT the flux distribution is almost homogeneous. When the field decreases, grey dendrites appear, see e. g. Fig.~\ref{fig:figure4}(e). The MOI signal associated with these dendrites is nonzero but lower than the signal seen in the center of the crystal. Upon further decreasing the field these dendrites grow in size and at $B_a=0$ they almost reach the center of the crystal. In addition, for fields below 0.8 mT the black dendrites appear, see Figs.~\ref{fig:figure4}(f)-(l). They grow towards the center of the crystal only for fields between 0.8 mT and 0.6 mT. For lower fields these dendrites change their shape but do not seem to grow any more. The black dendrites thus clearly behave differently from the dark ones.

\begin{figure}[ht]
	\centering
		\includegraphics[width=0.45\textwidth]{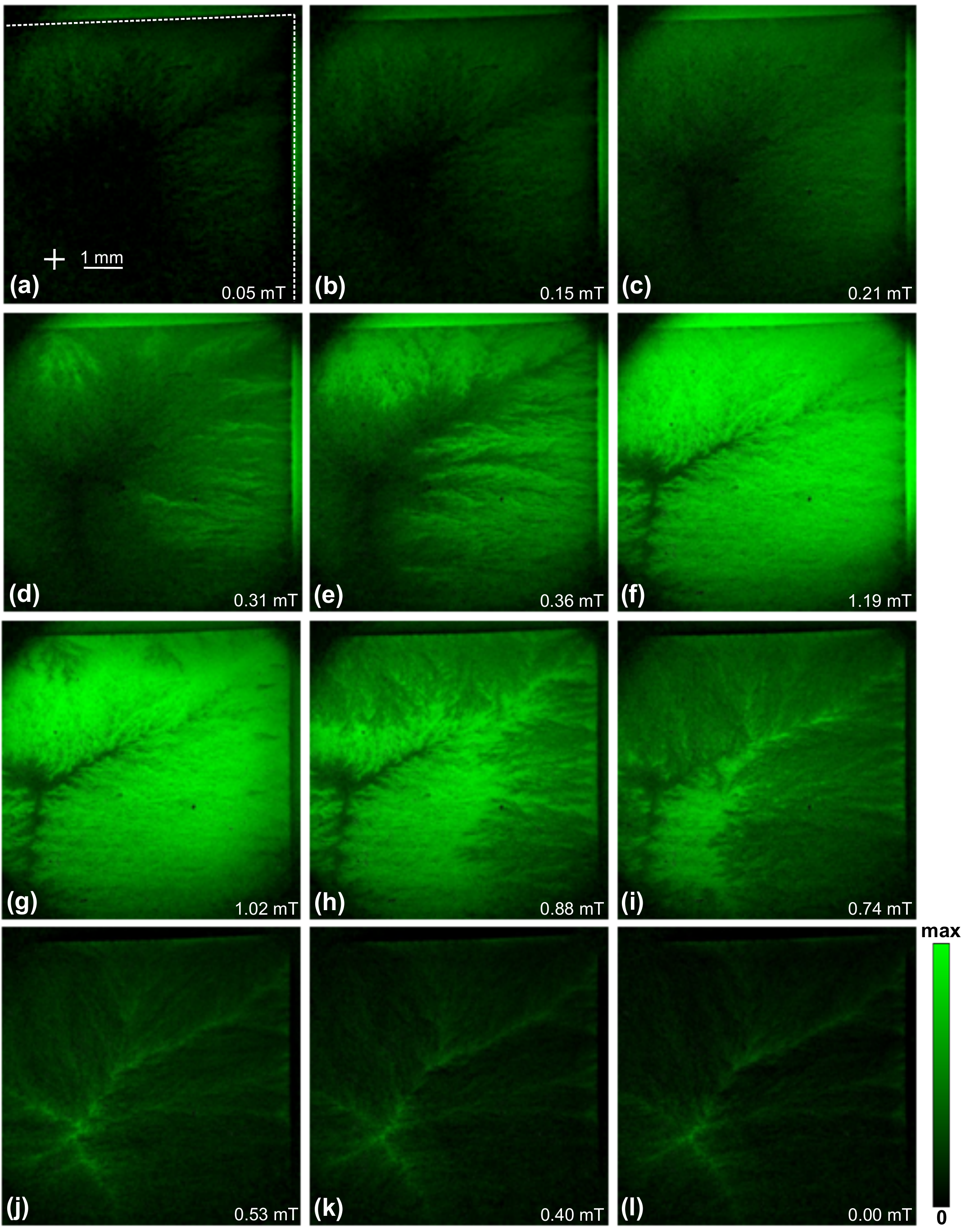}
	\caption{(Color online) MO images of dendritic flux penetration into bulk single crystal at $T/T_c \approx 0.996$. The applied magnetic field is oriented perpendicular to the sample and varied sinusoidally between 0 and 1.19 mT. Images (a)-(l) have been taken within the first period of the sweep. All images are subtracted from an image taken at $T/T_c \approx 0.996$ and $B_a=0$ taken prior to starting of the measurement. A video of the whole scan is shown in the supplemetary information (video Fig. 6). In (a) the sample edges are indicated by white dashed lines and the center of the crystal is indicated by a white cross.}
	\label{fig:figure5}
\end{figure}

Figure~\ref{fig:figure5} shows MOI images taken at $T/T_c \approx 0.996$, the closest value to $T_c$ where MOI data could be taken reliably. 
%
%%%%%%%%Figure%%%%%%%%%%%%%%%%%%%%%%%%%%%%%%%%%%%%%%%%%%%%%
%%%%%%%%%%%%%%%%%%%%%%%%%%%%%%%%%%%%%%%%%%%%%%%%%%%%%%%%%%%%%
Here, the applied field was increased from 0 to 1.19 mT and decreased back to 0. At 0.05 mT (Fig.~\ref{fig:figure5}(a)) some dendrites start penetrating the sample, producing a weak positive signal. These dendrites grow in size and, at 0.21 mT (Fig.~\ref{fig:figure5}(c)) almost reach the center. When the field is increased further to 0.31 mT, one observes additional dendrites producing a much brighter signal (Fig.~\ref{fig:figure5}(d)). These dendrites continue to grow towards the center of the sample up to the highest applied field of 1.19 mT (Fig.~\ref{fig:figure5}(f)). Note that these dendrites branch out more strongly than the dendrites that have entered at lower fields. With lowering the field (Figs.~\ref{fig:figure5}(g)-(l)) one first observes the appearance of dark dendrites (Fig.~\ref{fig:figure5}(g)) which again grow in size until they have almost reached the center. For fields below 0.5 mT black dendrites grow towards the center of the sample. Note, that black dendrites behave differently than their counterparts in Fig.~\ref{fig:figure4}, where they remain located near the edge of the crystal.

\section{Discussion}

\label{sec:Discussion}
The observed dendrites in the bulk Nb single crystal have features which are identical to those seen in thin films at low temperatures. Therefore, we can conclude that they are formed in a thin superconducting layer close to MOI indicator film. The appearance of a superconducting surface layer at temperature close to $T_c$ in a bulk sample is plausible \cite{StJames63, Casalbuoni05}. Surface superconductivity (SSC) occurs at magnetic fields $B_{c2}<B<B_{c3}$, where the critical surface field $B_{c3}$ for Nb is 1,695 times higher than upper critical field $B_{c2}$, at which superconductivity disappears in the bulk. The SSC persists in a surface sheath having a thickness which approximately equals the coherence length of the superconductor (about 40 nm for Nb at $T=0$, but significantly thicker at \textit{T} close to $T_c$). Near $T_c$, $B_{c2}$ as well as $B_{c3}$ strongly decrease and come into range of the fields used in MOI. The SSC is predicted for surfaces in a parallel field. Although our experiments are carried out in perpendicular field, they start from a ZFC state in which the magnetic field is initially expelled from the sample with the presence of a large field component parallel to the surface. At a fixed temperature, two sets of dendrites were observed. The first set shows a weak, and the second set strong MOI contrast. The situation could be therefore close to that described in \cite{Pan03, Pan04}, where above $B_{c2}$ the sample consists of three magnetically different layers: two SSC sheaths separated by a normal layer. By heating the crystal from below, bottom surface layer moves inside the sample and comes closer to the top surface superconducting layer. In-between there is a shrinking normal layer. The bottom layer is at a higher temperature and a bit farther from the indicator film, albeit sufficiently close to see flux changes by the MOI indicator film. Hence, one first observes dendrites in the bottom layer and they are in weak contrast. At slightly higher temperature dendrites penetrate the top superconducting layer and they are now close to the MOI indicator film and in strong contrast.

\section{Conclusions}

\label{sec:Conclusions}
In summary, we have shown the appearance of dendritic avalanches in a bulk Niobium single crystal in a very narrow temperature interval of about a tenth of a Kelvin near the critical temperature. The appearance of dendrites in bulk samples and at high temperatures is completely unexpected. In particular, we find that close to $T_c$ two sets of dendrites differing, at a fixed temperature, by the magnetic field required for their formation and by the maximum distance they penetrate into the sample. As avalanches could be employed in superconducting detectors \cite{Mikheenko13} and since temperatures close to $T_c$ are most important for these applications, our work is of significant practical importance. It also opens a route for the systematic study of dendritic avalanches, now also in bulk single crystals.
\\
\section*{Acknowledgements}

\label{sec:Acknowledgements}
The authors thank Rudolf H\"{u}bener for fruitful discussions.

\end{document}